\documentstyle[prl,aps]{revtex}
\begin{document}
\draft
\tighten
\onecolumn
\twocolumn[\hsize\textwidth\columnwidth\hsize\csname @twocolumnfalse\endcsname
\title{Phonon Squeezed States Generated
by Second Order Raman Scattering}
\author{Xuedong Hu$^1$ and Franco Nori$^2$}
\address{
1 Department of Physics, University of Illinois at Chicago, 845
W. Taylor, Chicago, IL 60607
\\
2 Department of Physics, The University of Michigan, Ann Arbor,
Michigan 48109-1120
}
\date{\today}
\maketitle
\begin{abstract}
We study squeezed states of phonons, which allow a reduction in 
the quantum fluctuations of the atomic displacements to below
the zero-point quantum noise level of coherent phonon states.
We investigate the generation of squeezed phonon states using a
second order Raman scattering process.
We calculate the expectation values and fluctuations of both the
atomic displacement and the lattice amplitude operators, as well as
the effects of the phonon squeezed states on macroscopically
measurable quantities, such as changes in the dielectric constant.
These results are compared with recent experiments.
\end{abstract}

\vspace*{-0.1in}
\vskip2pc]
\narrowtext

\vspace*{-0.1in}
{\it Introduction}.---Non-classical photon states 
such as various forms of squeezed states
have attracted much attention during the past decade \cite{special}.
These novel states are attractive because they have new statistical and
quantum mechanical properties.  For instance, some of these states can
achieve lower quantum noise than the zero-point fluctuations of the
vacuum or coherent states. Thus they provide a way of manipulating
quantum fluctuations and have a promising future in different
applications ranging from optical communications to gravitational wave
detection \cite{special}.  In recent years, squeezed states are also
being explored in a variety of non-quantum-optics systems, including
ion-motion and classical squeezing \cite{Rugar}, 
molecular vibrations \cite{Janszky}, 
polaritons \cite{Birman,xhuprb}, 
and phonons in crystals \cite{xhuaps,xhurmp,xhuprl}. 
Ref.~\cite{xhurmp,xhuprl} propose 
a second-order Raman scattering (SORS) process for phonon squeezing:
if the two incident light beams are in coherent states, 
the phonons generated by the SORS are in 
a two-mode squeezed state.  Here we consider both the continuous
wave case studied in \cite{xhurmp} and the impulsive 
case studied in \cite{Garrett}.  
The experimental realization of squeezed 
phonons \cite{Garrett} via a SORS process 
has brought attention to the subject of squeezed phonon states 
\cite{science}.

Regarding detection methods, \cite{xhurmp,xhuprl} proposed that 
if the first-order Raman scattering 
is either very weak or prohibited, the second-order stimulated 
Raman scattering process can be used to generate two-mode
phonon quadrature squeezed states.  Moreover, squeezed phonons
could be detected by measuring the intensity of the reflected 
probe light \cite{xhurmp,xhuprl}.  This method has been used to 
detect phonon amplitudes, since the reflectivity is 
closely related to the atomic displacements in a crystal.  
The same argument applies for the transmitivity.
Measuring a transmitted probe light pulse, Ref.~\cite{Garrett} 
observed squeezed phonons produced by an impulsive SORS.  
The intensity of the CW SORS signal for many materials 
might be too weak to be detected with current techniques,
but might be accessible in the future.

The SORS process originates from the quadratic term in 
the polarizability change $\delta \! P_{\alpha \beta}$.
The photon-phonon interaction $V$ that leads to the SORS
process is \cite{Born}
$V = - \frac{1}{4} \sum_{\alpha \beta} \sum_{\bf q}^N \sum_{j j'}
P_{\alpha \beta}^{{\bf q}j, -{\bf q}j'} Q_{{\bf q}j} Q_{-{\bf q}j'}
E_{1\alpha} E_{2\beta} \,$.
Here, $E_{1\alpha}$ and $E_{2\beta}$ are electric field amplitudes
along $\alpha$ and $\beta$ directions with frequencies $\omega_1$ and
$\omega_2$. 
The second-order polarizability tensor
$P_{\alpha \beta}^{{\bf q}j, -{\bf q}j'}$ 
satisfies 
$P_{\alpha \beta}^{{\bf q}j, -{\bf q}j'} = P_{\alpha \beta}^{-{\bf q}j',
{\bf q}j} = P_{\alpha \beta}^{-{\bf q}j, {\bf q}j'} \,$.
Recall that the complex normal mode operator $Q_{{\bf q}j}$ of the
phonons is related to the phonon creation $b_{-{\bf q}j}^{\dagger}$ and
annihilation $b_{{\bf q}j}$ operators by 
$Q_{{\bf q}j} = b_{{\bf q}j} + b_{-{\bf q}j}^{\dagger}$.
If the incident photon fields are not attenuated 
we can treat the optical fields as classical waves, and 
also consider the different pairs of $\pm {\bf q}$
modes as independent, and treat them separately.
Thus, for one particular pair of $\pm {\bf q}$ modes, the complete
Hamiltonian for the two phonon modes involved in the SORS
process has the form \cite{Born}:
$
{\cal H}_{\bf q}  =  H_{\bf q}
- \{ 4^{-1} \sum_{\alpha \beta} 
P_{\alpha \beta}^{{\bf q}, -{\bf q}} E_{1\alpha} E_{2\beta} \} 
Q_{\bf q} Q_{-\bf q} \; ,
$
\noindent where 
$
H_{\bf q} = \hbar \omega_{\bf q} 
\{ b_{\bf q}^{\dagger} b_{\bf q} + 
       b_{-\bf q}^{\dagger} b_{-\bf q} \}  
$
is the free phonon Hamiltonian for the modes ${\bf q}$ and $-{\bf q}$,
$\; \omega_{\bf q}= (\omega_1 - \omega_2 ) / 2 $, and the branch labels
$j$ and $j'$ have been dropped.

Here we consider two different cases.  The first is when the incident
photons are in two monochromatic beams \cite{xhurmp}; i.e., with 
electric fields 
$E_j = {\cal E}_j \cos({\omega_j t}+\phi_j)$; $j=1,\,2$.
In the second case the incident photons are in an ultrashort pulse whose 
duration is much shorter than the phonon period \cite{Garrett}.

{\it Squeezed Phonons via Continuous Wave SORS}.---Let 
us now first consider the continuous wave (CW) case.  Because the
photons are monochromatic, we can take a rotating wave approximation 
\cite{Schubert} and keep only the on-resonance terms in the Hamiltonian.
The off-resonance terms only contribute to virtual processes 
\cite{Srivastava} at higher orders.  This approximation is
appropriate for times much longer than the phonon period.  
The simplified Hamiltonian has the form
\begin{eqnarray}
{\cal H}_{\bf q}^{(cw)} & = & H_{\bf q}
- \lambda_{\bf q} \left\{ b_{\bf q} b_{-\bf q} \, 
e^{2i\omega_{\bf q} t + i\phi_{12}} + c.c.
\right\} \; , \nonumber \\
\lambda_{\bf q} & = & \frac{1}{16} | \sum_{\alpha \beta} P_{\alpha
\beta}^{{\bf q}, -{\bf q}} {\cal E}_{1\alpha} {\cal E}_{2\beta} | \,
,
\end{eqnarray}
where $\phi_{12}$ and $\lambda_{\bf q}$ refer to the overall phase and
amplitude, respectively, of the product of the 2nd-order polarizability
and the incident electric fields.  Recall that $P_{\alpha \beta}^{{\bf
q}, -{\bf q}}$ is real, therefore the phase $\phi_{12}$ has no ${\bf
q}$-dependence.  It originates solely from the two photon modes.
The Schr\"{o}dinger equation for the $\pm {\bf q}$--mode phonons is
$
i \hbar \partial_t \left|\psi_{\bf q} (t)\right\rangle
= {\cal H}_{\bf q}^{(cw)} (t) \left|\psi_{\bf q} (t)\right\rangle \, ,
$
and its time-evolution operator can be solved by a transformation into
the interaction picture.  The result can be expressed as \cite{xhurmp}
\begin{equation}
\left|\psi_{\bf q} (t) \right\rangle  =  
e^{ \{   H_{\bf q} \, t /  i\hbar  \} } \; 
e^{ \{ \zeta_{\bf q}^* b_{\bf q} b_{-\bf q}
- \zeta_{\bf q} b_{\bf q}^{\dagger} b_{-\bf q}^{\dagger} \} }
\left|\psi_{\bf q} (0) \right\rangle \;, \\
\label{eq:cwstate}
\end{equation}
where 
$\zeta_{\bf q}  =  - \; i \lambda_{\bf q} \, t \, e^{-i\phi_{12}} /
\hbar  \;$.  Notice that the second factor in the time-evolution
operator is a two-mode quadrature squeezing operator \cite{Loudon}.

In the CW case considered here, the amplitude of the squeezing factor
$\zeta_{\bf q}$ grows linearly with time.  However, this initial linear
growth will be eventually curbed by subsequent phonon-phonon scattering
and optical pump depletion.  In other words, the expression for the
squeezing factor $ \zeta_{\bf q} $ is valid for times much larger than
one phonon period, 
but much smaller
than phonon lifetimes (because this treatment considers non-decaying
phonons).  Indeed, if this growth rate is not fast enough compared to
the phonon decay rate, the squeezing effect may never reveal itself 
in an experiment.  In addition, the phase of the squeezing factor
is determined by the phase difference of the two incoming light waves.
If the $\pm {\bf q}$ phonon modes are initially in a vacuum state or in
a coherent state, the
SORS will drive them into a two-mode quadrature squeezed state
\cite{xhurmp}.

The time evolution operator of {\it all\/} the phonon mode pairs 
(instead of just one pair of $\pm {\bf q}$ modes) that are involved 
in this SORS process has the form 
$
U(t) = \prod_{\bf q} U_{\bf q}(t) \,
$.
Therefore, as long as the photon depletion is negligible, all the
phonon modes that are involved in a SORS 
process are driven into two-mode quadrature squeezed states.  In other
words, squeezing can be achieved in a continuum of phonon modes by a
CW stimulated SORS process.

{\it Squeezed Phonons via Impulsive SORS}.---Recently, an impulsive
SORS process has been used to experimentally generate phonon squeezing
\cite{Garrett}.  Here we treat the problem
expressing the time evolution operator of the system in terms of a
product of the two-mode quadrature squeezing operator and the free
rotation operators \cite{Schumaker}.  Since the incident photons are
now in an ultrashort pulse, the complete Hamiltonian
can be solved exactly in the limit when the optical field can be
represented by a $\delta$-function.  Such an approximation is usually
considered when the optical pulse duration is much shorter than the
optical phonon period, which is experimentally feasible with
femtosecond laser pulses.  The Hamiltonian for the SORS can now be
written as
$
{\cal H}^{'} 
=  \sum_{\bf q} \left\{ 
H_{\bf q}
- \lambda'_{\bf q} \delta(t) Q_{\bf q} Q_{-\bf q} \right\} \,$, 
where $\lambda'_{\bf q}$ carries the information on the amplitudes of
the incoming optical fields and the electronic polarizability.  Notice
that the light-phonon coupling strength $\lambda_{\bf q}$ in the CW
case has units of energy, while $\lambda'_{\bf q}$ here has units of
$\hbar$.  To further simplify the problem, we assume that only $\pm
{\bf q}$ modes are involved in the process.  Such a simplification is
possible when the photon depletion and the phonon anharmonic
interaction are negligible, so that different pairs of phonon modes are
independent from each other.  The Hamiltonian is now
\begin{equation}
{\cal H}_{\bf q}^{'} 
\, = \, H_{\bf q}
- \lambda'_{\bf q} \delta(t) Q_{\bf q} Q_{-\bf q} \,,
\label{eq:hpulse}
\end{equation}
and the Schr\"{o}dinger equation for these two phonon modes is
$
i \hbar \partial_t |\psi_{\bf q} (t)\rangle
\, = \, 
{\cal H}_{\bf q}^{'} 
\, |\psi_{\bf q} (t)\rangle 
$.
This equation can be solved by separating the free oscillator terms and
the two-phonon creation and annihilation terms. The resulting
time-dependent wavefunction is
\begin{eqnarray}
|\psi_{\bf q} (t)\rangle & = & 
\exp{ \left\{ \frac{ t H_{\bf q} }{ i \hbar } \right\} }
\ \exp{ \left\{ \frac{ i \lambda'_{\bf q} H_{\bf q} }
{ \hbar^2 \omega_{\bf q} } \right\} }  
\nonumber \\
& & \times \exp{ \left\{ \zeta_{\bf q}^{' \, *} b_{\bf q} b_{-\bf q} - 
\zeta'_{\bf q} b_{\bf q}^{\dagger} b_{-\bf q}^{\dagger} \right\} } 
|\psi_{\bf q} (0^-)\rangle \, .
\label{eq:implusestate}
\end{eqnarray}
Here $\zeta'_{\bf q} = -i\lambda'_{\bf q} \, e^{-i\lambda'_{\bf
q}/\hbar}/\hbar$.  Hence the effect of the optical pulse is clear:  it
first applies a two-mode quadrature squeezing operator on the initial
state, then rotates the state by changing its phase \cite{Schumaker}.
The state will then freely evolve after $t =0^+$.  This result is
consistent with Ref.~\cite{Garrett} where the time-evolution operator
is expressed in terms of real phonon normal mode operators \cite{Born},
instead of the complex ones used in this paper.  Notice that, in
contrast to the CW SORS, the phase of the squeezing factor $\zeta'$ for
the impulsive case is fixed by the intensity of the light pulse.

{\it Macroscopic Implications.}---Now that we have obtained the 
phonon states for both the CW and pulsed cases, let us consider 
the macroscopic implications of these states.
An experimentally observable quantity $O$ which is related to the atomic
displacements in the crystal can generally be expressed in terms of
$Q_{\bf q}$:
$ O = O(0) + \sum_{{\bf q}} (\partial O/\partial Q_{\bf q})
Q_{\bf q} + \ldots \, = O_0 + O_1 +  O_2 + \ldots $
where the first term $O_0 = O(0)$ is the operator $O$ when all $Q_{\bf
q}$'s vanish.  An example of an experimentally observable quantity $O$
is the change in the crystal dielectric constant $\delta \epsilon$ due
to the atomic displacement produced by the incident electric fields.
To first order in $Q_{\bf q}$, 
$
\delta \epsilon = \delta \epsilon_1 = \sum_{q_x > 0} \left| \frac{\partial
( \delta \epsilon )}{\partial Q_{\bf q}} \right|
\sqrt{\frac{\hbar}{2\omega_{\bf q}}} ( b_{{\bf q}} + b_{{-\bf
q}}^{\dagger} ) e^{i\Psi_{\bf q}} + ( b_{{-\bf q}} 
+ b_{{\bf q}}^{\dagger} ) e^{-i\Psi_{\bf q}} ] \,
$.
Here $\Psi_{\bf q}$ is the phase of $\partial O/\partial Q_{\bf q} =
\partial (\delta \epsilon)/\partial Q_{\bf q}$.  Indeed, a widely used
method to track the phases of coherent phonons in the time-domain
\cite{review} is based on the observation of the reflectivity (or
transmission) modulation $\delta R$ ($\delta T$) of the sample, which
is linearly related to $\delta \epsilon$---the change in the dielectric
constant due to lattice vibrations.
The above equation for $\delta \epsilon$ indicates that we can
define a generalized \cite{xhurmp} lattice amplitude operator
\cite{xhuprb,xhuprl}:
$ \
u_g (\pm {\bf q}) = \left( b_{\bf q} + b^{\dagger}_{-\bf q} \right)
e^{i\Psi_{\bf q}} + \left( b_{-\bf q} + b^{\dagger}_{\bf q} \right)
e^{-i\Psi_{\bf q}}  \,.
$
This generalized lattice amplitude $u_g (\pm {\bf q}) = 2 Re\{ Q_{\bf
q} e^{i\Psi_{\bf q}} \}$ is the underlying microscopic quantity related
to an observed reflectivity or transmission modulation when the linear
term in $Q_{\bf q}$, $\; \delta \epsilon_1$, exists.

Since different pairs of $\pm {\bf q}$ phonon modes are uncorrelated to
one another, the fluctuation of $O_1$($= \delta \epsilon_1$) can be
expressed as
$
\langle (\Delta \delta \epsilon_1)^2 \rangle = \sum_{q_x > 0} \left(
\hbar/2\omega_{\bf q} \right) 
$
$
\left| \partial \left( \delta \epsilon
\right) / \partial Q_{\bf q} \right|^2 \langle \Delta u_g^2 (\pm {\bf
q}) \rangle \,
$.
Here the state is $|\psi(t)\rangle = U(t) |\psi(0)\rangle = \prod_{\bf
q} U_{\bf q} (t) |\psi_{\bf q} (0)\rangle$ in either the CW or the
impulsive case.  We can again focus on a single pair of $\pm {\bf q}$
modes.  In the CW case, using Eq.~(\ref{eq:cwstate}), the fluctuation is
\begin{eqnarray}
\langle \Delta u_g^2 (\pm {\bf q}) \rangle^{(cw)}  & = & 2 \{
e^{-2r_{\bf q}} \cos^2 ( \Omega_{\bf q} (t) + \phi_{12}/2)
\nonumber \\  
& & + e^{ 2r_{\bf q}} \sin^2 ( \Omega_{\bf q} (t) + \phi_{12}/2 ) \} \, ,
\label{eq:cwfluc}
\end{eqnarray}
where 
$r_{\bf q} = |\zeta_{\bf q}| = \lambda_{\bf q} \; t /\hbar \,$, 
$\Omega_{\bf q} (t) = \omega_{\bf q} t + \pi / 4$,
and hereafter $\langle \ldots \rangle$ denotes an expectation value 
on squeezed states, unless stated otherwise.
Therefore, at certain times, the fluctuation $\langle \Delta u_g^2 (\pm
{\bf q}) \rangle^{(cw)}$ can be smaller than $2$, which is the vacuum
fluctuation level.  Furthermore, all the pair of phonon modes that are
driven by the stimulated SORS 
process share the same frequency: 
$\omega_{\bf q} = (\omega_1 - \omega_2)/2$.  Therefore, all
the fluctuations $\langle \Delta u_g^2 (\pm {\bf q}) \rangle^{(cw)}$
evolve with the same $\omega_{\bf q}$.  Notice that there is no
dependence on $\Psi_{\bf q}$ in the final expression of $\langle \Delta
u_g^2 (\pm {\bf q}) \rangle^{(cw)}$, and the squeezing factor phase
$\phi_{12}/2$ has no ${\bf q}$--dependence, all the pairs of modes
involved through the SORS share the same phase in their fluctuations.
Therefore there can be squeezing in the overall fluctuation $\langle
(\Delta \delta \epsilon_1)^2 \rangle^{(cw)}$.  Furthermore, the phase of
this overall fluctuation can be adjusted by tuning the phase difference
of the two incoming light beams.

In the impulsive case \cite{Garrett,review}, if the $\pm {\bf q}$-mode
phonons are driven into a squeezed vacuum state, the fluctuation in
$u_g (\pm {\bf q})$ is
\begin{eqnarray}
\langle \Delta u_g^2 (\pm {\bf q}) \rangle'
= 2 \{ e^{-2r'_{\bf q}} \cos^2 \Omega'_{\bf q} (t) +
e^{2r'_{\bf q}} \sin^2 \Omega'_{\bf q} (t) \} \,,
\label{eq:impulsefluc} 
\end{eqnarray}
where $r'_{\bf q} = |\zeta'_{\bf q}| = \lambda'_{\bf q} /\hbar \, ,$
and $\Omega'_{\bf q} (t) = \Omega_{\bf q} (t) - r'_{\bf q} $.
Again, the squeezing will reveal itself through oscillations in
$\langle [\Delta (\delta \epsilon_1)]^2 ({\bf q}) \rangle'$ which is
proportional to $ \langle \Delta u_g^2 (\pm {\bf q}) \rangle'$.  Note
that these oscillations are essentially the same as the ones obtained 
in the CW case.  However, now the squeezing factor is time-independent.
Also, the $t=0$
phase $\pi/4 - r'_{\bf q}$ in Eq.~(\ref{eq:impulsefluc}) 
is ${\bf q}$--dependent.
Eq.(\ref{eq:impulsefluc}) can be rewritten as 
$
\langle \Delta u_g^2 (\pm {\bf q}) \rangle' = 2 \left\{ \cosh 2r'_{\bf q} 
+ \sinh 2r'_{\bf q} \; \sin ( 2\omega_{\bf q} t - r'_{\bf q} ) \right\} \,. 
$
For small $r'_{\bf q}$, this becomes 
$\langle \Delta u_g^2 (\pm {\bf q}) \rangle' = 2 \{ 1 + 2 r^{' \, 
2}_{\bf q} + 2 r'_{\bf q} \sin ( 2\omega_{\bf q} t - r'_{\bf q} ) \}$.
This expression has essentially the same form as the one 
obtained in \cite{Garrett}: 
$\langle Q_{\bf q}^2 (t) \rangle = 
 \langle Q_{\bf q}^2 (0) \rangle 
\{ 1 + 2 \xi_{\bf q}^2 + 2 \xi_{\bf q} 
\sin( 2 \omega_{\bf q} t + \varphi_{\bf q}) \}$.  
The small phase term $\varphi_{\bf q}$ is neglected 
in \cite{Garrett} when computing transmission changes.
The difference in phases, 
$r'_{\bf q}$ versus $\varphi_{\bf q}$,
% which is of second order in $r'_{\bf q}$, 
is negligible in the limit of very small squeezing factor, and
originates from the different interaction Hamiltonians used here and in
\cite{Garrett}.  The interaction term in \cite{Garrett} is proportional
to $u_g^2(\pm {\bf q})$ with $\Psi_{\bf q} = 0$ (notice that their
$Q_{\bf q}$ is real and based on standing wave quantization
\cite{Born}).  Therefore, the interaction Hamiltonian in \cite{Garrett}
is (in our notation)
$
V \propto u_g^2(\pm {\bf q}) \propto 2 Q_{\bf q} Q_{-\bf q} + Q_{\bf
q}^2 + Q_{-\bf q}^2 \,
$.
However, the last two terms in this expression do not satisfy momentum
conservation, we thus did not include them and kept only $Q_{\bf q}
Q_{-\bf q}$ in our interaction term (this form is also used by
Ref.\cite{Born}).

If the linear perturbation $\delta \epsilon_1$ due to phonons is
negligible, such as in \cite{Garrett}, then the second order correction
$O_2 (= \delta \epsilon_2)$
must be considered.  When the phonon states are modulated by a SORS, so
that the $\pm {\bf q}$ modes are the only ones which are correlated,
then 
$
\delta \epsilon_2 = \sum_{{\bf q}} \frac{\partial^2 (\delta \epsilon) }
{\partial Q_{\bf q} \, \partial Q_{-\bf q}} Q_{\bf q} Q_{-\bf q} \,.
$
Let us first focus on one pair of $\pm {\bf q}$ modes in the CW case.
In a vacuum state,
$
\langle 0 | Q_{\bf q} Q_{-\bf q} | 0 \rangle = 1 \,
$;
while in a squeezed vacuum state $|0 \rangle_{\rm sq} $,
$
{ }_{\rm sq}\langle 0 | Q_{\bf q} Q_{-\bf q} | 0 \rangle_{\rm sq} 
= \langle \Delta u_g^2 (\pm {\bf q}) \rangle^{(cw)}/2 \,
$,
with the right hand side given in Eq.~(\ref{eq:cwfluc}).
Therefore, the expectation value of $Q_{\bf q} Q_{-\bf q}$ in a
squeezed vacuum state is periodically smaller than its vacuum state
value.  Let us now include all the phonon modes that contribute to
$\delta \epsilon_2$.  In a vacuum state,
$
\langle 0| \delta \epsilon_2 |0 \rangle = \sum_{\bf q} \partial^2
\delta \epsilon/(\partial Q_{\bf q} \, \partial Q_{-\bf q}) \,
$.
On the other hand, in a squeezed vacuum state,
\begin{equation}
\langle \delta \epsilon_2 \rangle = \frac{1}{2}\sum_{{\bf q}}
\frac{\partial^2 (\delta \epsilon)}{\partial Q_{\bf q} \, \partial
Q_{-\bf q}} \langle \Delta u_g^2 (\pm {\bf q}) \rangle^{(cw)} \,.
\end{equation}
\noindent Since the phase $\phi_{12}/2$ has no ${\bf q}$--dependence,
contributions from the phonon modes sharing the same frequency add up
constructively.  It is thus possible that 
$\langle \delta \epsilon_2 \rangle$ is periodically smaller than its
vacuum state value. 
Similarly, in the impulsive case, 
$
\langle \delta \epsilon_2 \rangle' = 2^{-1} \sum_{{\bf q}}
\frac{\partial^2 (\delta \epsilon)}{\partial Q_{\bf q} \, \partial
Q_{-\bf q}} \langle \Delta u_g^2 (\pm {\bf q}) \rangle' \, ;
$
However, the phase factor in $\langle \delta \epsilon_2 \rangle'$ 
has a ${\bf q}$--dependence through $r'_{\bf q}$, so that all the 
phonon modes with the same $\omega_{\bf q}$ do {\it not\/} contribute 
to $\langle \delta \epsilon_2 \rangle'$ synchronously.  
In the CW SORS and in the very-small-$r'_{\bf q}$ limit 
impulsive SORS the phase of the expectation value 
$\langle Q_{\bf q} Q_{-\bf q}\rangle$ does not depend on ${\bf q}$;
this is crucial to the experimental observation of modulations 
in the dielectric constant, because this ${\bf q}$--insensitivity 
leads to constructive summations of all the ${\bf q}$ pairs involved.  
Also, at a van Hove singularity a large number of modes contribute 
to $\delta \epsilon_2$ with the same frequency and phase,
thus their effect is larger and easier to observe \cite{Garrett}.

{\it Squeezed Phonons via a finite-width SORS.---}Of course, real 
light pulses are not $\delta$-functions. Therefore, we have also 
considered a SORS pumped by a light pulse with a finite width
(smaller than the phonon period $T$) 
instead of a $\delta$-function. 
For a fixed peak height $I$, we find\cite{more} 
that the optimal pulse width $T_{\rm p}^{\rm opt}$ 
that maximizes the squeezing effect satisfies 
$T_{\rm p}^{\rm opt} \approx T / 4.4 $.  
This calculation indicates that the experiments\cite{Garrett} 
used a pulse width which is nearby the optimal value 
($T / 4.4 \approx 300 / 4.4 $ fs $\approx 68$ fs $ \approx T_{\rm p}$).  
The calculation\cite{more} can be summarized as follows.
First, in the impulsive Hamiltonian we replace the 
$\delta$-function by a Gaussian with its 
width $T_{\rm p}$ as a variational parameter.  
Since now the Hamiltonian is time-dependent in the interaction picture, 
we cannot directly integrate the Schr\"{o}dinger equation.  Instead, 
we use the Magnus method 
to obtain the time evolution operator and keep only the dominant 
first term.  This approximation is valid when the pulse duration is 
shorter than the phonon period.  We then calculate the width 
$T_{\rm p}^{\rm opt} $  
of the Gaussian that maximizes the squeezing factor.
For a constant peak intensity, 
a pulse that is too narrow does not contain enough photons; while
it can be proven that a pulse which is too long 
(i.e., with a width comparable to $T$), 
attenuates the squeezing effect.

{\it Phonon Squeezing Mechanism.}---What is the mechanism of phonon 
squeezing in the SORS processes?  For the CW case,  
the Hamiltonian is the same as an
optical two-mode parametric process\cite{Schubert}, with the low frequency
interference of the combined photon modes as the pump, the two phonon
modes as the signal and idler.  The frequencies of these modes satisfy
$\omega_{\bf q} + \omega_{-\bf q} = 
\omega_{1} - \omega_{2}$.  
The impulsive case is slightly different.  Although the
Hamiltonian is similar to a parametric process, the energy transfer
from the photons to the two phonon modes is instantaneous.  The
resulting phonon state is a two-mode quadrature squeezed vacuum state.
Indeed, a regular parametric process pumps energy into the signal and
idler modes gradually, while the impulsive SORS does it suddenly.
The correlation between the two phonon modes, and thus the squeezing
effect, is also introduced instantaneously.  Notice that this mechanism
is reminiscent of the frequency-jump mechanism proposed in
\cite{Janszky}.  In the impulsive SORS, the frequency of the phonon
modes has an ``infinite'' $\delta$-peak change at $t = 0$, while the
frequency-jump mechanism has finite frequency changes, and squeezing
there can be intensified by repeated frequency jumps at appropriate
times.  However, as it has been pointed out in \cite{Janszky}, a finite
frequency jump up immediately followed by an equal jump down results in
no squeezing at all.

In conclusion, we have studied theoretically the generation of phonon
squeezing using a stimulated SORS process.  In particular, we
calculated the time evolution operators of the phonons in two different
cases: when the incident photons are in monochromatic continuous waves,
and when they are in an ultrashort pulse.  The amplitude of the
squeezing factor initially increases with time and then saturates in
the CW SORS case, while it remains constant in the pulsed SORS case.
In addition, the $t = 0$ phase
of the squeezing factor in the CW SORS, $\phi_{12}$, 
can be continuously adjusted by tuning the relative 
phase of the two incoming monochromatic photon beams, 
while for the pulsed SORS the phase 
($\propto \lambda_{\bf q}^{\prime}$) of the squeezing factor 
is determined by the amplitude of the incoming light pulse
% (although this phase factor does not contribute to the 
% fluctuations).
%
%
For both cases we calculated 
the quantum fluctuations of a generalized lattice amplitude 
operator and 
the second order contribution to the change in dielectric 
constant, which is measurable.
For the finite-width impulsive case, we computed the optimal pulse
width, in terms of the phonon period, that maximizes the squeezing 
effect.

We acknowledge useful conversations with S.~Hebboul, R.~Merlin,
S.~Tamura and H.~Wang.  One of us (XH) acknowledges support from the US
Army Research Office.

\end{document}